\documentclass[twocolumn]{emulateapj}
\usepackage{txfonts}
\usepackage{natbib}
\usepackage{graphicx}
\usepackage{amssymb}
\bibpunct{(}{)}{;}{a}{}{,} 

\shorttitle{Testing Bulge Extinction Maps with APOGEE}
\shortauthors{Schultheis et al.}

\begin{document}
\title{Extinction Maps towards the Milky Way Bulge: 2D and 3D Tests with APOGEE}
\author{M.~Schultheis\altaffilmark{1},
 G.~Zasowski\altaffilmark{2,3}, \\
 C.~Allende Prieto\altaffilmark{4,5},
 F.~Anders\altaffilmark{6,7},
 R.~L.~Beaton\altaffilmark{8},
 T.~C.~Beers\altaffilmark{9,10},
 D.~Bizyaev\altaffilmark{11,12},
 C.~Chiappini\altaffilmark{6,13},
 P.~M.~Frinchaboy\altaffilmark{14}
 A.~E.~Garc{\'i}a P\'erez\altaffilmark{8},
 J.~Ge\altaffilmark{15},
 F.~Hearty\altaffilmark{16},
 J.~Holtzman\altaffilmark{12},
 S.~R.~Majewski\altaffilmark{8},
 D.~Muna\altaffilmark{17},
 D.~Nidever\altaffilmark{18},
 M.~Shetrone\altaffilmark{19},
 D.~P.~Schneider\altaffilmark{16,20}
}

\altaffiltext{1}{Universit\'e de Nice Sophia-Antipolis, CNRS, Observatoire de C\^ote d'Azur, Laboratoire Lagrange, 06304 Nice Cedex 4, France; mathias.schultheis@oca.eu}
\altaffiltext{2}{Department of Physics \& Astronomy, Johns Hopkins University, Baltimore, MD 21218, USA; gail.zasowski@gmail.com} 
\altaffiltext{3}{NSF Astronomy \& Astrophysics Postdoctoral Fellow}
\altaffiltext{4}{Instituto de Astrof\'{\i}sica de Canarias, Calle V{\'i}a L{\'a}ctea s/n, E-38205 La Laguna, Tenerife, Spain} 
\altaffiltext{5}{Departamento de Astrof{\'i}sica, Universidad de La Laguna, E-38206 La Laguna, Tenerife, Spain} 
\altaffiltext{6}{Leibniz-Institut f\"ur Astrophysik Potsdam (AIP), 14482 Potsdam, Germany}	
\altaffiltext{7}{Technische Universit\"at Dresden, Institut f\"ur Kern- und Teilchen-physik, 01069 Dresden, Germany} 
\altaffiltext{8}{Department of Astronomy, University of Virginia, Charlottesville, VA 22904, USA} 
\altaffiltext{9}{National Optical Astronomy Observatory, Tucson, AZ 85719, USA} 
\altaffiltext{10}{Joint Institute for Nuclear Astrophysics (JINA), Michigan State University, East Lansing, MI 48824, USA} 
\altaffiltext{11}{Apache Point Observatory, Sunspot, NM 88349, USA} 
\altaffiltext{12}{New Mexico State University, Las Cruces, NM 88003, USA} 
\altaffiltext{13}{Laborat\'orio Interinstitucional de e-Astronomia, - LIneA, Rio de Janeiro, RJ - 20921-400, Brazil} 
\altaffiltext{14}{Department of Physics \& Astronomy, Texas Christian University, TCU Box 298840, Fort Worth, TX 76129, USA} 
\altaffiltext{15}{Astronomy Department, University of Florida, Gainesville, FL 32611, USA} 
\altaffiltext{16}{Department of Astronomy and Astrophysics, The Pennsylvania State University, University Park, PA 16802} 
\altaffiltext{17}{Department of Astronomy, The Ohio State University, Columbus, OH 43210, USA} 
\altaffiltext{18}{Department of Astronomy, University of Michigan, Ann Arbor, MI 48109, USA} 
\altaffiltext{19}{McDonald Observatory, The University of Texas at Austin, Austin, TX 78712, USA} 
\altaffiltext{20}{Institute for Gravitation and the Cosmos, The Pennsylvania State University, University Park, PA 16802} 

\begin{abstract}
{Galactic interstellar extinction maps are  powerful and necessary tools for Milky Way structure and stellar population analyses, 
particularly toward the heavily-reddened bulge and in the midplane. 
However, due to the difficulty of obtaining reliable extinction measures and distances for a large number of stars that are independent of these maps,
tests of their accuracy and systematics have been limited.}
 {Our goal is to assess a variety of photometric stellar extinction estimates, including both 2D and 3D extinction maps, using independent extinction measures based
 on a large spectroscopic sample of stars towards the  Milky Way bulge. }
{We employ stellar atmospheric parameters derived from high-resolution $H$-band APOGEE spectra, combined with theoretical stellar isochrones, to calculate
line-of-sight extinction and distances for a sample of more than 2400 giants towards the Milky Way bulge.  We compare these extinction values to those
predicted by individual near-IR and near+mid-IR stellar colors, 2D bulge extinction maps  and 3D  extinction maps.}
{The long baseline, near+mid-IR stellar colors are, on average, the most accurate predictors of the APOGEE extinction estimates,
and the 2D and 3D extinction maps derived from different stellar populations along different sightlines show varying degrees of reliability.  
We present the results of all of the comparisons and discuss reasons for the observed discrepancies.
We also demonstrate how the particular stellar atmospheric models adopted can have a strong impact on this type of analysis, 
and discuss related caveats.}
 \end{abstract}

\setcounter{footnote}{0}

\section{Introduction} \label{sec:intro}
Interstellar extinction remains one of the primary obstacles to studying the structure and stellar populations
of the Galactic bulge.
Some of the earliest bulge 
 extinction maps were made using optical photometry, primarily of red clump (RC) stars, 
from microlensing surveys such as OGLE and MACHO 
(e.g., \citealt{stanek1996}; \citealt{sumi2004}; more recently \citealt{kunder2008}; \citealt{nataf2013}).
 With the arrival of large area, near-infrared (IR) photometric surveys such as DENIS and 2MASS, additional extinction maps became available.  
For example, \citet{schultheis1999} and \citet{dutra2003} used red giant branch (RGB) stars, together with stellar evolutionary models, to trace extinction up to $A(V) \sim 25$~mag with a spatial resolution of 4\arcmin.
\citet{gosling2008} used the near-IR colors of bulge stars to trace small scale interstellar dust variations ($\sim$\,5\arcsec) in the Galactic Center. 
\citet{gonzalez2012} also used RC stars to trace the interstellar dust extinction based upon data from the VVV survey,
which reaches sufficiently faint  magnitudes  to use the RC population even in the most highly extinguished regions, such as the Galactic Center. 
However, these extinction maps are two-dimensional, 
and when applying them, one implicitly assumes that all stars are located at a distance beyond the typical distance probed by the map. 
At the level of individual stars, \citet{majewski2011} introduced the ``Rayleigh-Jeans Color Excess'' (RJCE) method, and 
demonstrated that near- to mid-IR colors (e.g., $H-4.5\mu$m)
 could be used to measure the effects of the interstellar dust on a star-by-star basis, largely independent of stellar type.

Only a few 3D extinction maps towards the Galactic bulge area have been constructed thus far. 
\citet{drimmel2003} built a theoretical 3D Galactic dust distribution model,
based on the interstellar dust and stellar distribution inferred from the COBE near- and far-IR emission.
\citet{marshall2006} provided a 3D dust extinction map by comparing 
2MASS data with the Besan\c{c}on stellar population synthesis model (\citealt{robin2003}). 
An improved version of the Marshall et al. map, using VVV and GLIMPSE-II data and an updated version of the Besan\c{c}on model (\citealt{robin2012}), 
has been published by \citet{chen2012}; this was later expanded by \citet{schultheis2014} for the full VVV bulge area.

In this paper, we use stellar properties derived from new, high-resolution, near-IR spectra 
 to probe the inner Milky Way's interstellar extinction in three dimensions, and  compare the results
with existing 2D and 3D extinction maps. This allows, for the first time, a detailed comparison
of the derived extinction and distances with the available 3D maps in the literature. 
In Section~\ref{sec:sample}, we outline the sample of stars used in this study, 
and in Section~\ref{sec:extdistances}, we describe the derivation of their extinctions and distances.
In Sections~\ref{sec:compare2d} and \ref{sec:compare3d}, 
we compare these extinctions to existing 2D and 3D extinction maps, respectively,
and present explanations for the discrepancies,
including a discussion of systematic differences as a function of stellar parameters.
Finally, in Section~\ref{sec:discussion}, we discuss the impact of the choice of stellar models and extinction law
on our results.

\section{The Sample} \label{sec:sample}

\subsection{APOGEE}
One of four experiments in the Sloan Digital Sky Survey III (SDSS-III; \citealt{eisenstein2011}),  
 the Apache Point Observatory Galactic Evolution Experiment (APOGEE; \citealt{majewski2010}) is a large scale, near-IR, 
high-resolution ($R \sim 22\,500$) spectroscopic survey of Milky Way 
stellar populations. 
The survey uses a dedicated, 300-fiber, cryogenic spectrograph coupled to the wide-field, Sloan 2.5\,m telescope (\citealt{gunn2006}) at Apache Point Observatory (APO).
APOGEE observes in the $H$-band ($1.5\,\mu{\rm m} - 1.7\,\mu{\rm m}$), 
where extinction by dust is significantly lower than at optical wavelengths 
(e.g., $A(H) / A(V) \sim 0.16$). APOGEE observes, at high signal-to-noise ratio (${\rm S/N} \sim 100$ per Nyquist-sampled pixel), 
about 100\,000 red giant stars selected from the 2MASS survey, down to a typical flux limit of $H \sim 12-14$ (\citealt{zasowski2013}). Approximately 85\% of  our bulge stars have a  magnitude brighter than $\rm H < 11$, and all have $\rm H < 12.2$.


Stars are observed using standard SDSS plug-plates, which normally have a field of view (FOV) radius of $1.5^\circ$,
but the high airmass of the bulge observed from APO (${\rm latitude} \sim 32^\circ$) produces strong differential refraction effects on stars near
the plug-plate edges.  Therefore, APOGEE's bulge fields have stars no more than $0.9^\circ$ from the
field center (some fields observed early in the survey are even smaller, with $R < 0.5^\circ$) to mitigate this effect.

With its high resolution and high S/N, 
APOGEE will determine both accurate radial velocities (to better than $\rm 0.5\,km\,s^{-1}$ external accuracy) and
precise abundance measurements for most of the vast stellar sample, 
including the most abundant metals in the universe (C, N, O),
along with other $\alpha$, odd-$Z$, and iron-peak elements. 
The latest SDSS-III Data Release (DR10; \citealt{ahn2013}) 
provides spectra of about 55\,000 stars to the scientific community, as well
as the derived stellar properties, including radial velocities, effective temperatures, surface gravities, and metallicities. 
Additional information, such as photometry and target selection criteria, is also provided and described in \citet{zasowski2013}.

\subsection{Stellar Parameters}

Stellar parameters 
are determined by the APOGEE Stellar Parameters and Chemical Abundances Pipeline 
\citep[ASPCAP;][and in prep]{garciaperez2014}. 
These values are based on a $\chi^{2}$-minimization between observed and synthetic model spectra
performed with the {\sc FERRE} code \citep[][and subsequent updates]{allende2006}.
Model spectra are interpolated on a regular grid computed with the ASS$\epsilon$T code \citep{koesterke2008,koesterke2009},
a custom line list specially compiled for the survey (M.~Shetrone et al., in prep),
and \cite{castelli2004} model atmospheres.
New ATLAS9 model atmospheres computed by \citet{meszaros2012} with varying C and $\alpha$ content,
relative to the solar composition from \citet{asplund2005}, will be used in future data releases. 

The accuracy of the DR10 ASPCAP $T_{\rm eff}$, $\log{g}$, and [M/H] values was evaluated by \citet{meszaros2013}.
Using a sample of well-studied field and cluster stars, including a large number of stars with asteroseismic stellar parameters from NASA's {\it Kepler} mission (\citealt{borucki2010}),
they compared ASPCAP results to the literature values.
They conclude that the  ASPCAP temperatures agree with other spectroscopic temperatures from the literature, 
with a mean offset of only 8~K and a 1$\sigma$ scatter of 161~K. 
 For literature photometric temperatures derived with the Infrared Flux Method (\citealt{gonzalez2009}),
 larger systematic differences were found, and a correction function was provided to convert the
ASPCAP temperatures to photometrically-calibrated temperatures. 
In the present work, we adopt the raw ASPCAP spectroscopic temperatures because these estimates,
based on continuum-normalized spectra, are independent of the interstellar extinction.

ASPCAP surface gravities are, in general about 0.2--0.3\,dex  larger than both isochrone and seismic
values in the range $\rm -0.5 < [M/H] < +0.1$, with increasing offsets at lower metallicities. 
An empirical correction has been calculated for  use in our analysis. Metallicities agree 
 with literature values for $\rm -0.5 < [M/H] < +0.1$ (within 0.1\,dex), but at both the metal-poor and
 metal-rich end, systematic offsets of up to 0.2--0.3\,dex are apparent. Again, a correction factor has been
derived, which is applied here.

In summary, we adopt for our analysis the spectroscopic temperatures (not corrected) from ASPCAP and
apply the correction terms given in \citet{meszaros2013} for the ASPCAP surface gravities and metallicities.

\begin{figure}[!htbp]
   \includegraphics[width=9.0cm]{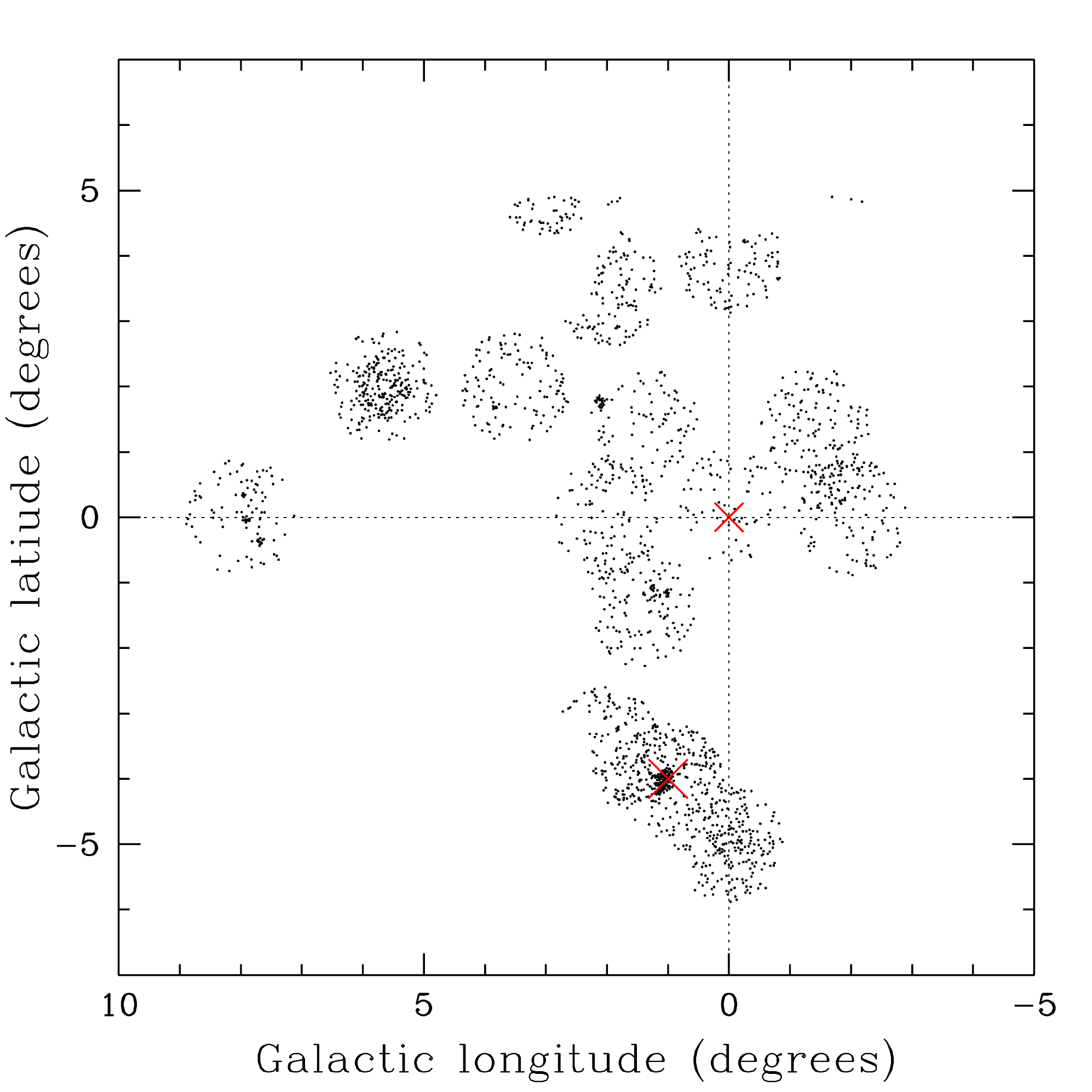}
\caption{Galactic longitude and latitude of our final APOGEE sample towards the Galactic bulge. 
The positions of the Galactic Center and Baade's Window are marked with crosses. 
}
\label{gallong}
\end{figure}

\subsection{Sample Used In This Work}
The initial selection for our sample comprises all  APOGEE targets from the first two years of the survey that are located towards the Galactic bulge:
$ -10^\circ < l < 10^\circ$ and  $-10^\circ < b < 5^\circ$, coinciding with the footprint of the VVV survey extinction map (\citealt{gonzalez2012}). 
These include data that are part of  DR10 (comprising 60\% of the sample), along with data not included in that release.  For all stars,
 we use parameters from the v400 version of APOGEE's combined reduction + analysis pipeline.

We then filtered the stars to remove those with parameters close to the edges of the model grid, as described by \citet{meszaros2013}. 
We selected stars with $\rm S/N > 50$, $\chi^{2}_{\rm ASPCAP} < 30$, and $\log{g} < 3.5$.
to ensure a sample of stars with reliable ASPCAP fits and  stellar parameters. 
No additional selection criteria in  $T_{\rm eff}$ or [M/H] have been applied, though we did reject stars where these values were not well-matched by the  stellar models described below (Sect.~\ref{sec:extdistances}). Figure~\ref{gallong} shows the distribution in Galactic longitude ($l$) and Galactic latitude ($b$)
of our final sample of 2433 stars. 
  

\begin{figure}[!htbp]
\includegraphics[angle=0,width=0.5\textwidth]{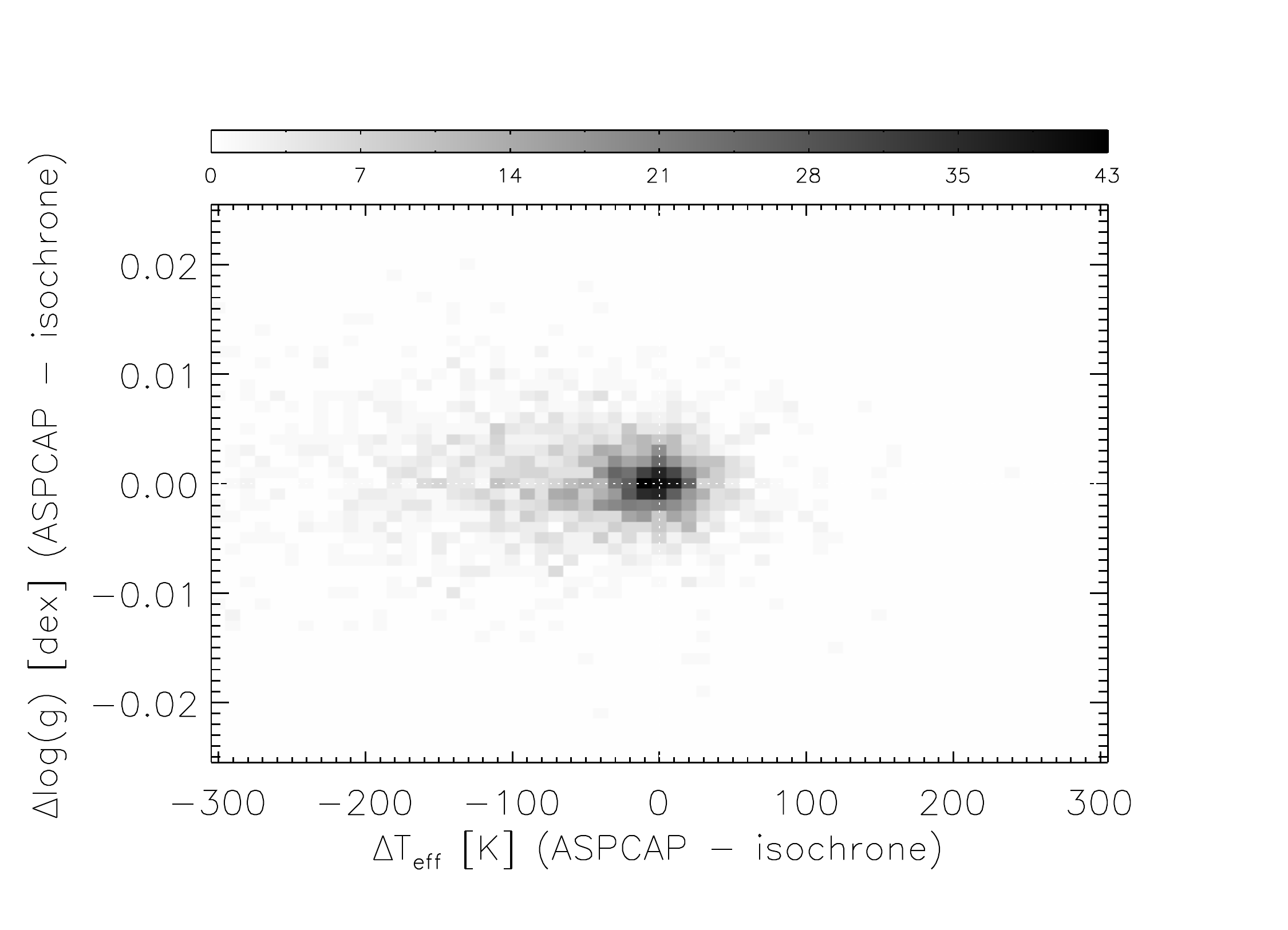}
\caption{Difference in $T_{\rm eff}$ and $\log{g}$ (ASPCAP -- ``isochrone'') of our sample. 
The peak in both distributions is at zero, although due to the irregular grid in $T_{\rm eff}$ and $\log{g}$ of the Padova isochrones, the tail of the $\Delta T_{\rm eff}$
distribution is asymmetric towards more negative values.}
\label{DeltaTeffdeltalogg}
\end{figure}

\section{Derivation of Extinction and Distances} \label{sec:extdistances}

We used the Padova isochrones set\footnote{http://stev.oapd.inaf.it/cgi-bin/cmd} from \citet{marigo2008}  along with the \citet{girardi2010} Case A correction for 
low-mass, low-metallicity AGB tracks that were  matched with the ASPCAP parameters $T_{\rm eff}$,
$\log{g}$, and [M/H], for our sample of APOGEE targets. 
Our isochrone grid has
metallicity steps of 0.2~dex between $\rm -2.5 < [Fe/H] < +0.5$ and 
age steps of $\rm \Delta(log\,age)=0.05\,Gyr$. 
The intrinsic Padova model grid is slightly irregular in $T_{\rm eff}$ and $\log{g}$ depending on the $T_{\rm eff}$ range, but the steps in mass are sufficiently small to ensure
a typical resolution better than 100\,K in $T_{\rm eff}$ and 0.1\,dex in $\log{g}$. However, we note that the isochrone grid samples very poorly the parameter space  within $\rm 3500 < T_{\rm eff} < 4000\,K$ and $\rm  log{g} > 2.0$.
 ASPCAP estimates [M/H], not [Fe/H],  using multiple elements. 
As pointed out by  \citet{meszaros2013}, [M/H] is, in general close to [Fe/H], particularly after the \citet{meszaros2013} calibration is applied --- within 0.1\,dex or so. 
We therefore assume that the [Fe/H] values of the isochrones are equivalent to [M/H]. For alpha-enhanced  stars  this relation might be not valid, and  could therefore introduce additional errors in the distance and extinction determination.

For each star, we  selected the isochrone closest in metallicity and then identified
the closest point in the corresponding $\rm T_{eff}$ and $\log{g}$ plane of the isochrone. No interpolation has been done, 
but stars that are too far from a point in the isochrone-grid ($|\Delta T_{\rm eff}| > 500$~K or
$|\Delta \log{g}| > 0.5$~dex) are discarded. Figure~\ref{DeltaTeffdeltalogg} shows the differences in $T_{\rm eff}$ and $\log{g}$ between the spectroscopic values
and those of the best-matched Padova isochrone point for each star (in the sense of ASPCAP -- ``isochrone''). 
While the differences in $\log{g}$ are rather small ($< 0.025\,{\rm dex}$), 
the effective temperatures can differ by several hundred K, mainly due to the irregular grid spacing in the 
$T_{\rm eff}$ vs. $\log{g}$ plane of the Padova isochrones. No dependency on
metallicity was found. The assymetric tail seen in Fig.~\ref{DeltaTeffdeltalogg} is composed of  stars that fall in the
parameter space $\rm 3500 < T_{eff} < 4000\,K$ and $\rm  log{g} > 2.0$, not covered
 by the isochrone grid. These stars comprise  $\sim$3\% of the sample. As a secondary effect, $\alpha$-enhanced bulge stars might not be as well represented by the assumption of [M/H] = [Fe/H] and solar-scaled isochrones.

These $\log{g}$ and $T_{\rm eff}$ offsets are incorporated into the uncertainties in the final extinction and distance estimates (see below).

\begin{figure}[!htbp]
   \includegraphics[width=9.0cm]{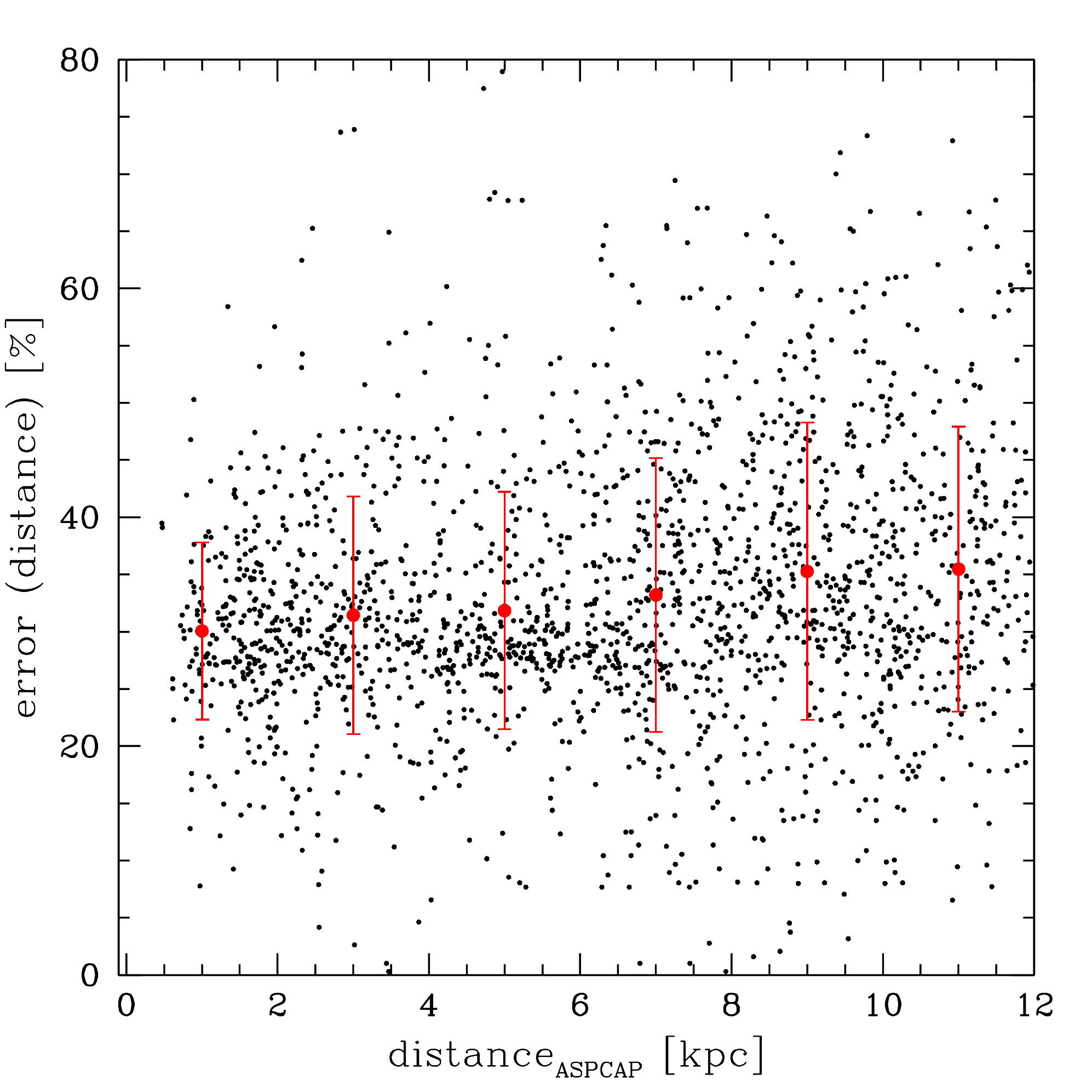}
\caption{Fractional distance error as a function of distance. The error in the distance includes both the
error in the isochrone-matching method and the typical errors in the stellar parameters derived by ASPCAP. The red points show the median values and the error bars the standard deviation in the six distance bins.
}
\label{distancevserrordist}
\end{figure}

Each star in the APOGEE sample has 2MASS magnitudes $J$, $H$, and $K_s$.
In the corresponding isochrone grid, we find the absolute magnitudes $ M_{J}$, $ M_{H}$ and $ M_{Ks}$. 
The color excess $E(J-K_s)$ can then simply be calculated by  $E(J-K_s) = J-K_s - (M_{J}-M_{Ks})$, 
where $J$ and $K_s$ are the star's observed 2MASS magnitudes and $ (M_{J}-M_{Ks})$  
the intrinsic, unreddened color from the isochrones. To convert $E(J-K_s)$ to $A(K_s)$, one has to assume a certain extinction law. 
We use here the relationships of \citet{nishiyama2009}, with $A(K_s) = 0.528 \times E(J-K_s)$,
which was derived for sightlines near the Galactic Center.
The extinction derived using this method is hereafter referred to as $A(K_s)_{\rm ASPCAP}$. Note that $A(K_s)_{\rm ASPCAP}$ is not a product of ASPCAP. 
The extinction law towards the Galactic bulge remains somewhat uncertain, although it has been shown to  
vary along different lines of sight (e.g., \citealt{gao2009}; \citealt{fritz2011}; \citealt{chen2012}; \citealt{nataf2013}).
 In Section~\ref{sec:discussion}  we examine the potential impact of these variations on our findings.

The  distance between each star and the isochrone grid  in the  $\Delta  T_{\rm eff}$ 
and $\Delta \log{g}$ dimensions (see Figure~\ref{DeltaTeffdeltalogg}),
 together with the individual errors $\sigma_{\rm Teff}$ and $\sigma_{\log{g}}$ from the ASPCAP pipeline
 \citep[those derived empirically by][]{meszaros2013}, 
give the total error in $T_{\rm eff}$ and $\log{g}$ for each star:

$\rm err_{Teff} = \sqrt{\Delta T_{\rm eff}^2 + \sigma_{\rm Teff}^2}$ and
$\rm err_{log\,g} = \sqrt{(\Delta \log{g})^2 + \sigma_{\rm log\,g}^2}$.

\noindent For each star, we added these uncertainties to the ASPCAP values (i.e., $T_{\rm eff} \pm {\rm err_{Teff}}$ and $\log{g} \pm  {\rm err_{log\,g}}$) 
and redid the isochrone matching, thus estimating the typical uncertainty 
in $A(K_s)_{\rm ASPCAP}$ for this method. 
However, these errors do not include systematic contributions from the choice of the stellar atmosphere models, isochrones, etc.
We discuss this issue in Section~\ref{stellib}.

Our distances were calculated using:
\begin{equation}
d = 10 ^{0.2 (K_s - M_{K_s}) + 5 - A(K_s)_{\rm ASPCAP}},
\end{equation}

\noindent with $K_s$ the 2MASS apparent magnitude, $M_{Ks}$  the absolute $K_s$ magnitude from the Padova isochrone, and the extinction
$A(K_s)_{\rm ASPCAP} $  as described above. The errors in the distance are obtained in the same way as for $A(K_s)_{\rm ASPCAP}$, including the errors in $T_{\rm eff}$ and $\log{g}$.
  

The resulting median errors in our derived distances from the isochrone-matching method, including the ASPCAP errors in $T_{\rm eff}$ and $\log{g}$, are on the order
of $\sim$30\%--40\% (see Fig.~\ref{distancevserrordist}). We compared our distances with those of \cite{anders2014}, 
who use a more sophisticated Bayesian approach based on \citet{allende2008} to compute SDSS distances both for APOGEE giants and SEGUE dwarfs. In general, there is  good agreement between the two distance scales. For
small heliocentric distances ($d < 2$~kpc) we find smaller values than Anders et al., by about 20\%, whereas for larger distances ($d > 6$~kpc) we tend to find  slightly larger values.
The r.m.s scatter between our work and  Anders et al. is about 30\%, similar to
the typical intrinsic error of our distances (Figure~\ref{distancevserrordist}).
In Table 1, we present the derived extinctions  (and associated distances) for our sample, along with the extinction values based on the literature 2D maps as described in the next section.

\begin{table*}[!htbp]
\caption{Derived extinction and distances from our APOGEE sources. The full catalog contains 2433 stars and is only available in electronic form.}
\begin{tabular}{lllllllllll}
Object&R.A.&DEC&$A(K_s)$&$\sigma(A(K_s)$&dist&$\sigma(dist)$&$A(K_s)_{\rm RJCE}$)&$A(K_s)_{\rm N12}$ &$A(K_s)_{\rm G12}$&$A(K_s)_{\rm EHK}$)\\
  &$\deg$&$\deg$&mag&mag&kpc&\%&mag&mag&mag&mag\\
\hline
2M17515147-2215539&  267.9644470&  -22.2649879&   0.218&   0.034&   7.449&  59.189&   0.213&   0.229 &  0.230&   0.328\\
2M17515740-2229440&  267.9891663&  -22.4955711&   0.171&   0.024&   2.316&  31.137&   0.203&   0.312&   0.261&   0.158\\
2M17515917-2221365&  267.9965515&  -22.3601379&   0.238&   0.037&   5.140&  34.498&   0.254&   0.263&   0.244&   0.344\\
2M17520342-2326376&  268.0142822&  -23.4438000&  0.290&   0.029&   4.626&  39.296&   0.369&   0.345&   0.331&   0.404\\
2M17520525-2248283&  268.0218811&  -22.8078842&   0.281&   0.038&   6.936&  46.518&   0.297&   0.288&   0.250&   0.487\\
\hline
\end{tabular}
\end{table*}
\section{Comparison to Individual Stellar Extinctions and 2D Extinction Maps} \label{sec:compare2d}

\subsection{Stellar Extinction Estimates and 2D Maps Used}
We compare our isochrone-based extinctions,  $A(K_s)_{\rm ASPCAP}$, to the following data:
\begin{itemize}
\item The individual stellar RJCE extinction estimates, $A(K_s)_{\rm RJCE}$, following the method of \citet{majewski2011}.  
We explored using both the {\it Spitzer}-IRAC 4.5$\mu$m and the WISE W2 (4.6$\mu$m) filters (together with 2MASS $H$),
but the larger pixel size of WISE is particularly disadvantageous in the crowded bulge, so we opt
to use IRAC data exclusively for $A(K_s)_{\rm RJCE}$.
\item The individual stellar extinction estimates    derived from the $E(H-K_s)$  color excess, $A(K_s)_{\rm EHK}$, following \citet{lada1994}, and assuming that all stars share a common intrinsic $(H-K_s)_0 = 0.13 $ color.  This assumption is only the first step of the fuller NICE extinction-mapping method
(e.g., \citealt{lada1994}; \citealt{lombardi2001}; \citealt{gosling2009}),
but statistically-cleaned NICE maps of large bulge regions have not yet been constructed. 
\item The extinction map based on RC stars by \citet[][hereafter G12]{gonzalez2012}. Using the BEAM
 calculator webpage\footnote{http://mill.astro.puc.cl/BEAM/calculator.php}, 
 we retrieved for each star the extinction in the map's 2$^\prime$ pixel
closest to the star's position, $A(K_s)_{\rm G12}$.
\item The ``all stars, median'' extinction map using the RJCE method by \citet[][hereafter N12]{nidever2012}. 
With the query scripts provided in that paper, we retrieved, for each star, the extinction in the 2$^\prime$ pixel
closest to the star's position, $A(K_s)_{\rm N12}$.

\end{itemize}

\begin{figure}[!htbp]
   \includegraphics[width=9.0cm]{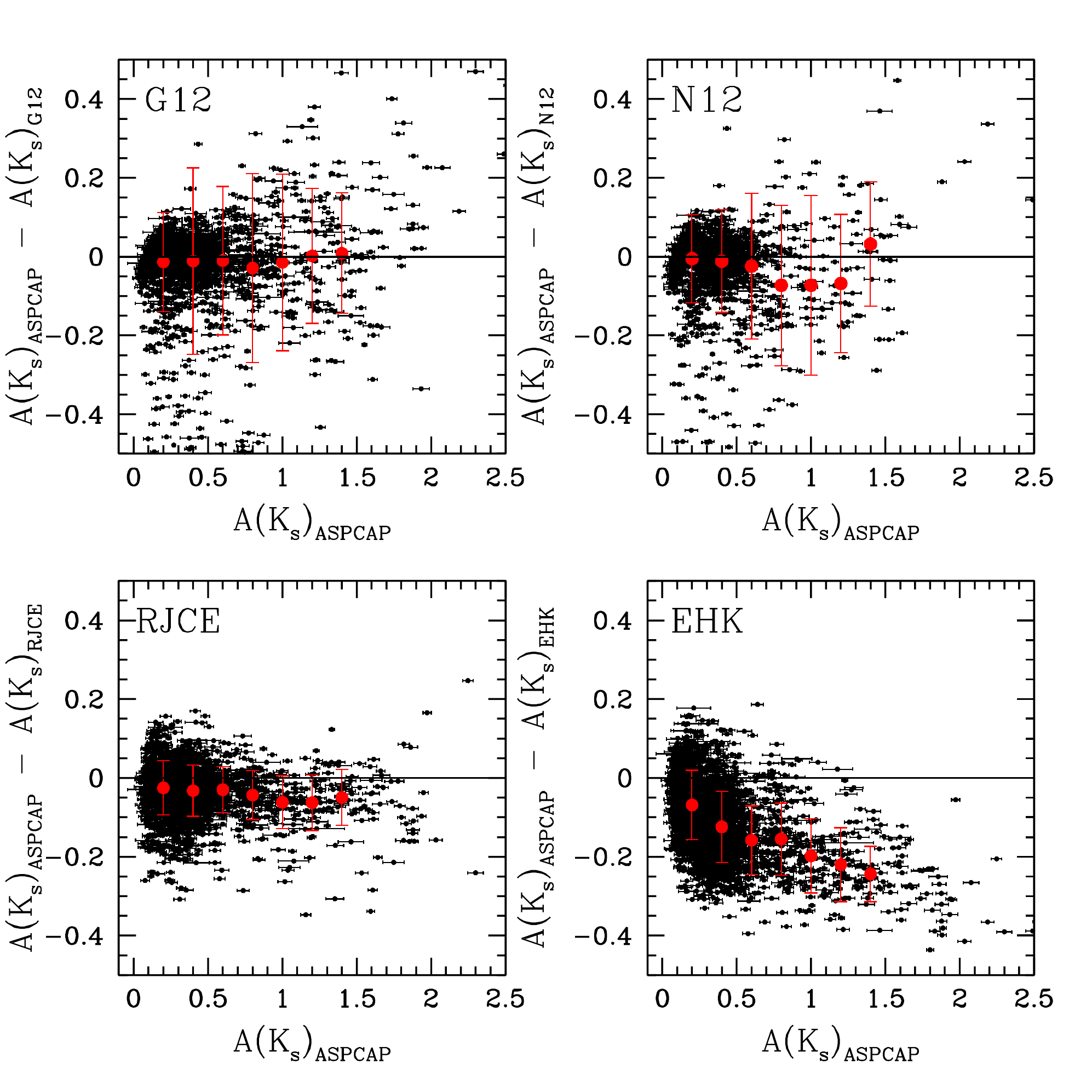}
\caption{Comparison of our derived extinction, $A(K_s)_{\rm ASPCAP}$,  to those of G12 (upper left), 
N12 (upper right), RJCE (lower left) and EHK (lower right), as a function of ASPCAP extinction. The error bars are the results from the isochrone
matching (see text). The red points show the median values and dispersions  
in $A(K_s)_{\rm ASPCAP}$ bins spaced 0.2 mag apart.}
\label{AkvsdeltaAk}
\end{figure}

\begin{figure}[!htbp]
   \includegraphics[width=9.0cm]{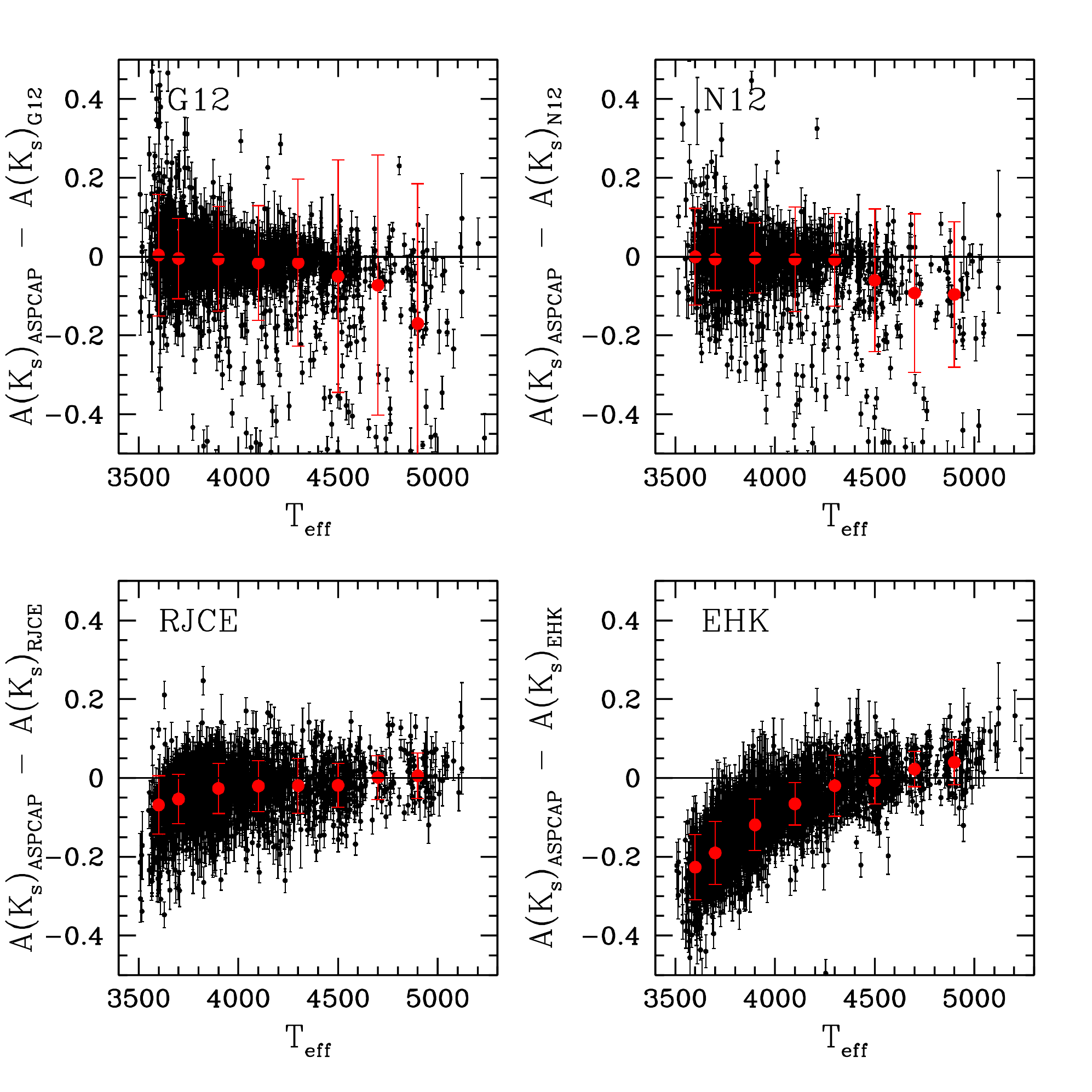}
\caption{Comparison of our derived extinction, $A(K_s)_{\rm ASPCAP}$, to those of G12 (upper left), N12 (upper right), RJCE (lower left) and EHK (lower right), as a function of $\rm T_{eff}$. The error bars are the results from the isochrone
matching (see text). The red points show the median value and dispersion.}
\label{TeffvsdeltaAk}
\end{figure}

\begin{figure}[!htbp]
   \includegraphics[width=9.0cm]{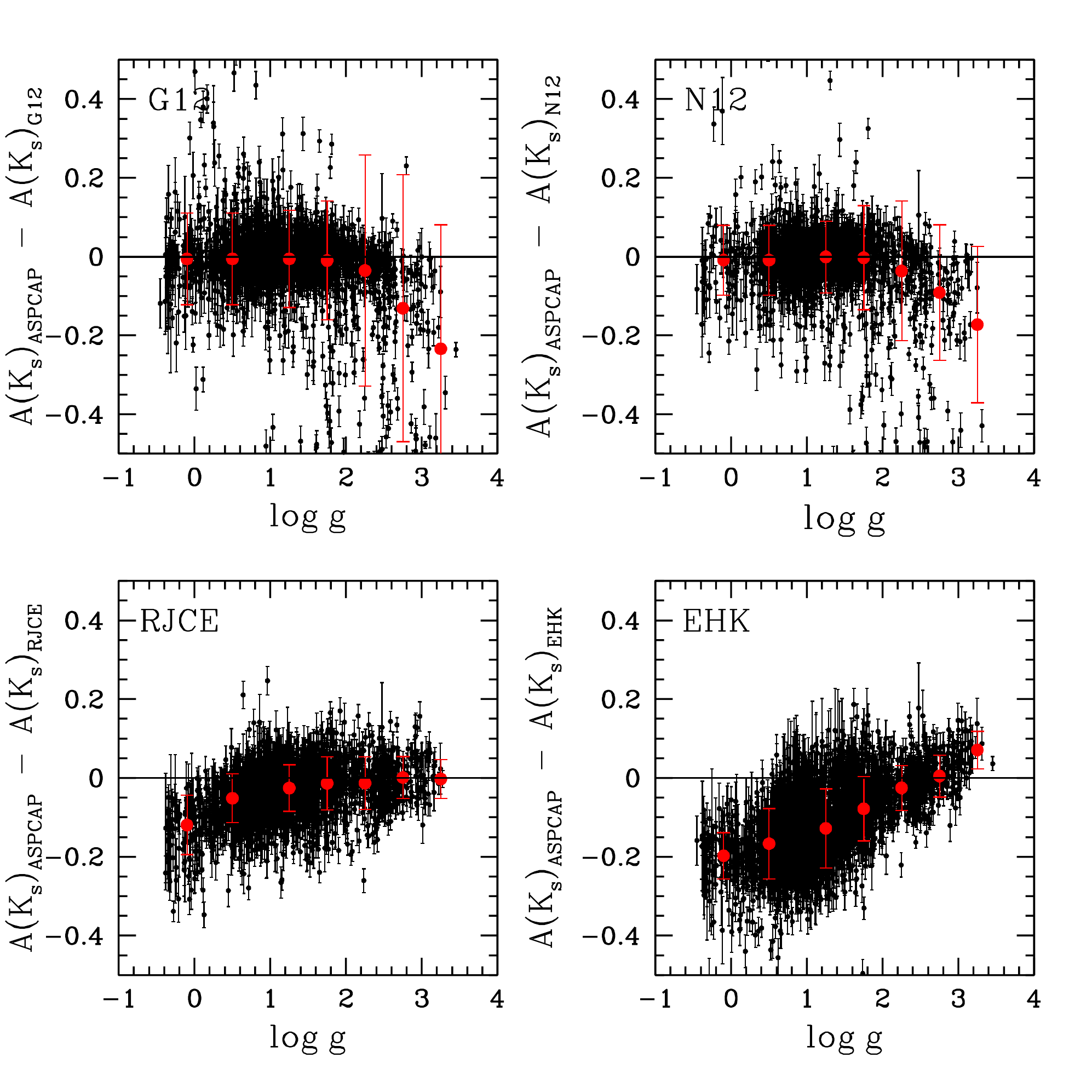}
\caption{Comparison of our derived extinction, $A(K_s)_{\rm ASPCAP}$, to those of G12 (upper left), N12 (upper right), RJCE (lower left) and EHK (lower right), as a function of log\,g. The error bars are the results from the isochrone
matching (see text). The red points show the median value and dispersion.}
\label{loggvsdeltaAk}
\end{figure}

\begin{figure}[!htbp]
   \includegraphics[width=9.0cm]{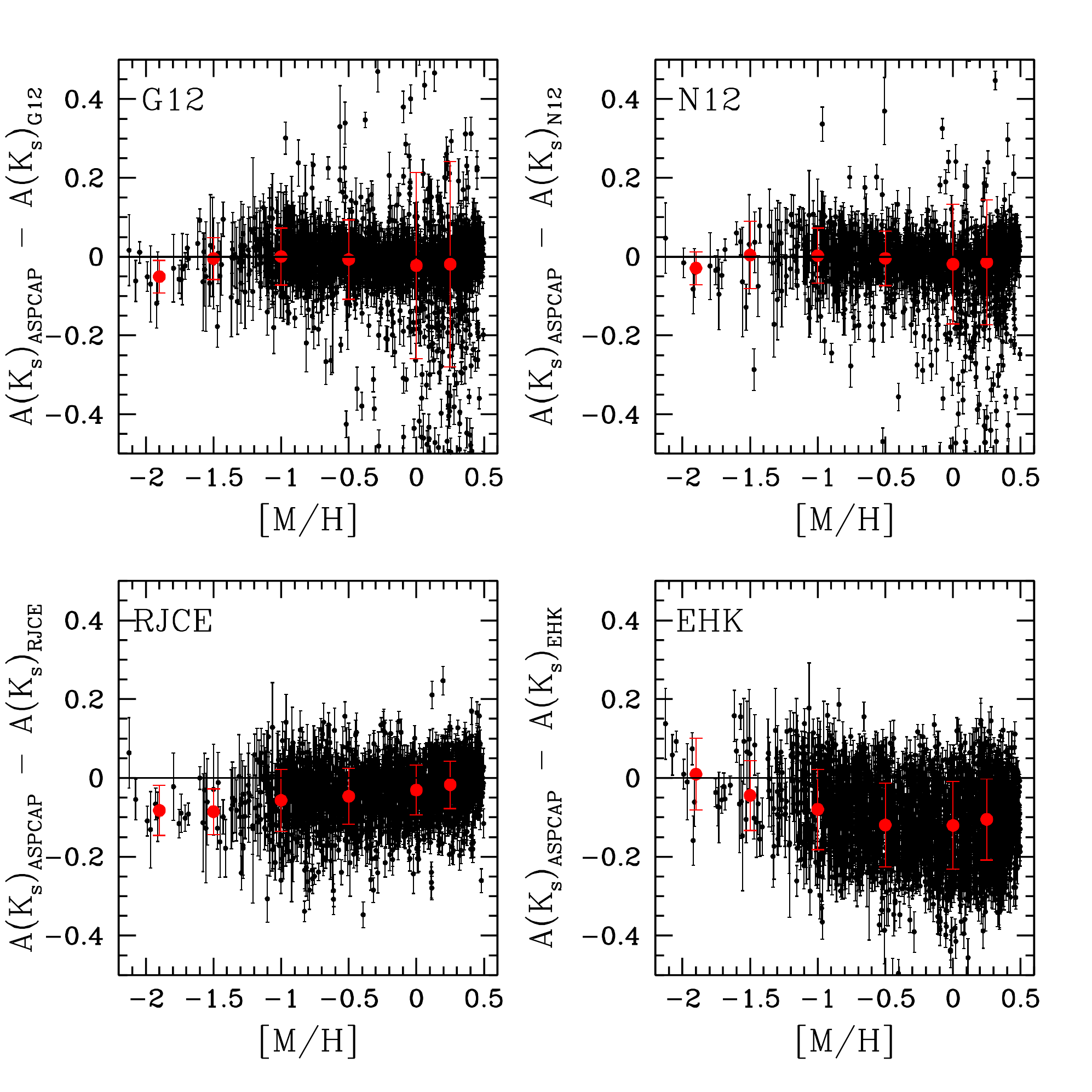}
\caption{Comparison of our derived extinction, $A(K_s)_{\rm ASPCAP}$, to those of G12 (upper left), N12 (upper right), RJCE (lower left) and EHK (lower right), as a function of [M/H]. The error bars are the results from the isochrone
matching (see text). The red points show the median value and dispersion.}
\label{FeHvsdeltaAk}
\end{figure}

All reddening or extinction values were transformed, if necessary, to $A(K_s)$ 
with the extinction law of \citet{nishiyama2009}. 

\subsection{Results and Discussion} \label{sec:discussion2d}

Figure \ref{AkvsdeltaAk} shows the comparison of our derived extinction, $A(K_s)_{\rm ASPCAP}$,  
with: G12 in the upper left panel,
N12 in the upper right panel, RJCE in the lower left panel, and  EHK in the lower right panel. 
We include for each star the typical error resulting from the isochrone matching. Each panel does not include the same number of stars, due  to the fact that  some stars lack  reliable extinction measurements from all four methods because of their inhomogeneous photometry. These stars ``missing'' one or more estimates have no preferred $A(K_s)_{\rm ASPCAP}$, and represent $\sim$20\% of the sample.

Overall, the   three  methods (G12, N12, RJCE) agree fairly well,
within $\rm |\Delta A(K_s)| < 0.2$~mag, while EHK is clearly discrepant. The smallest dispersion relative to this work is seen in comparison to the
RJCE method, where even at larger $A(K_s)$ the difference from $A(K_s)_{\rm ASPCAP}$ is smaller than 
0.1 mag (compared to a typical $A(K_s)_{\rm ASPCAP}$ uncertainty of $\sim$0.05 mag). 
The differences from the G12 maps are centered on zero but have a higher dispersion, with large differences at higher
$A(K_s)_{\rm ASPCAP}$ and a set of disparate points with $A(K_s)_{\rm ASPCAP} \lesssim 0.5$. The N12 map predicts similar extinctions as the RJCE values
for $A(K_s)_{\rm ASPCAP} < 1$, but the differences become more scattered for larger $A(K_s)$, similar to those found for G12.
The EHK method consistently overestimates the extinction, increasing at higher $A(K_s)_{\rm ASPCAP}$.

Figures~\ref{TeffvsdeltaAk}, \ref{loggvsdeltaAk}, and \ref{FeHvsdeltaAk}
show the extinction offsets as functions of the stellar $T_{\rm eff}$, $\log{g}$, and [M/H], respectively.
While for RJCE only a slight trend with $T_{\rm eff}$ is observed, on the order of the $A(K_s)_{\rm ASPCAP}$ uncertainty for $T_{\rm eff} \lesssim 3800$~K, 
the EHK method systematically overestimates $A(K_s)$ starting at cooler temperatures ($T_{\rm eff} < 4200$~K).
N12 and G12 overestimate $A(K_s)$, compared to ASPCAP, when $T_{\rm eff} > 4500$~K. 
With respect to surface gravity (Figure~\ref{loggvsdeltaAk}), EHK systematically overestimates extinction
for $\log{g} < 2$ and G12 and N12 for $\log{g} > 2.5$, respectively. 
RJCE overestimates $A(K_s)$ for the most luminous stars with $\log{g} < 1$.
For $\rm [M/H] < -1$, the RJCE extinction measurements
 deviate (Figure~\ref{FeHvsdeltaAk}), indicating that the assumption of a constant $(H-4.5\mu{\rm m})_0$ color
 used in the RJCE estimates is not valid
for low metallicities (as also noted in \citealt{zasowski2013}), while EHK actually overestimates extinction
 at higher [M/H]; 
 however, we are limited by the poor statistics for those kind of stars.

In summary, the G12 and N12 maps behave similarly in over-predicting extinction for main-sequence dwarfs and red clump or RGB stars with low reddening.
Most likely this effect is due to the fact that these maps are heavily weighted by the mean total extinction along the line of sight towards
the specific stellar tracers used to make the maps.  In both cases, these tracers are stars located in the bulge itself (RC giants for G12
and RGB giants for N12), so any low-reddening foreground sources will be overcorrected by these particular 2D maps (Section~\ref{sec:2d-vs-3d-maps}).

\begin{figure*}[!htbp]
\includegraphics[bb= 0 20 430 580,width=14cm,angle=90]{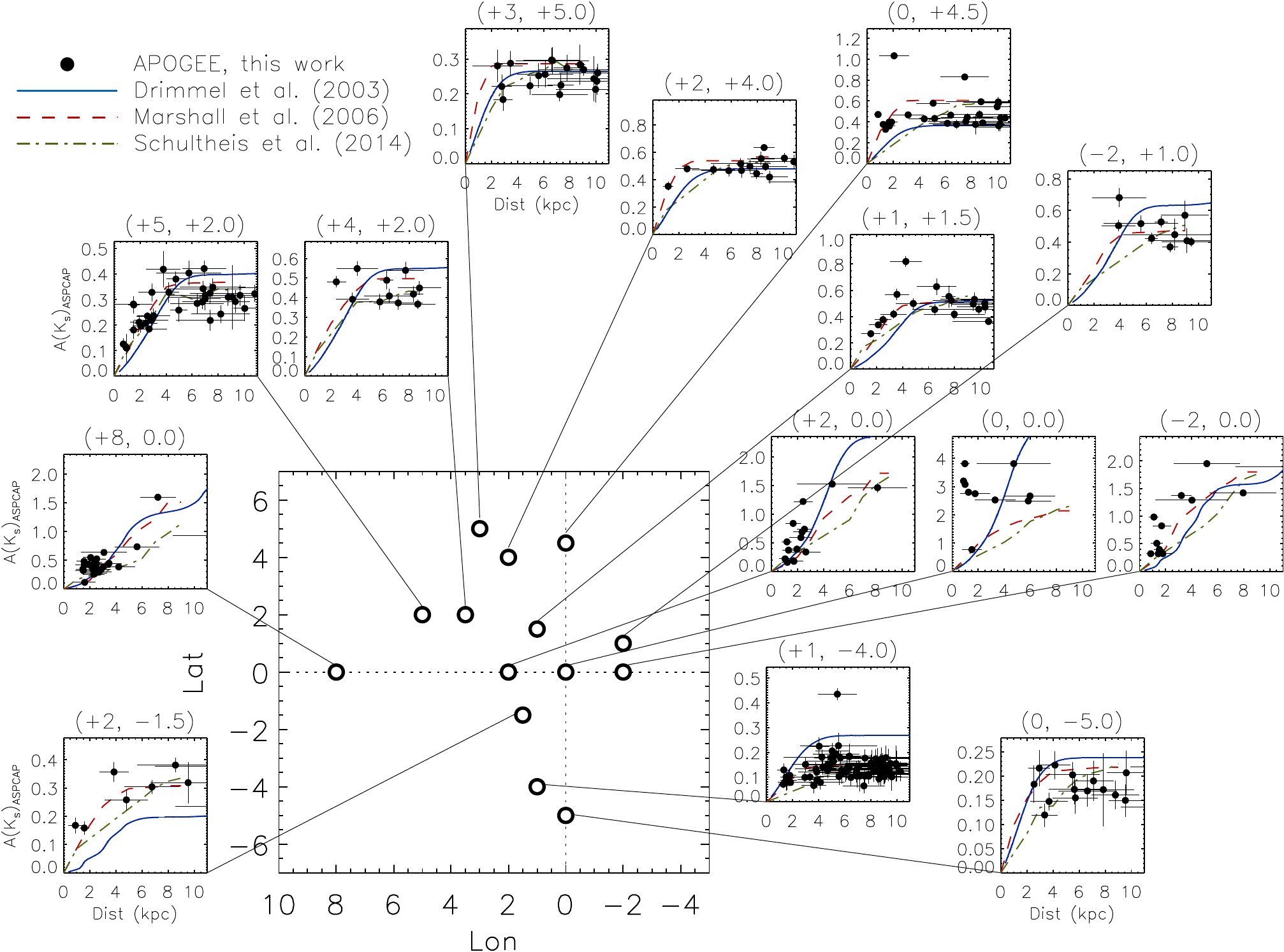}
\caption{$A(K_s)_{\rm ASPCAP}$, as a function of heliocentric distance (black points), compared 
to the 3D interstellar extinction maps of \citet[][solid blue line]{drimmel2003}, \citet[][dashed red line]{marshall2006}, 
and \citet[][dash-dot green line]{schultheis2014} for multiple lines of sight.}

\label{los1}
\end{figure*}

Extinctions derived from individual stellar color excesses do not suffer from this problem, since they only measure the impact
of dust along the line of sight to each star.  In the cases of $A(K_s)_{\rm RJCE}$ and $A(K_s)_{\rm EHK}$, 
offsets from $A(K_s)_{\rm ASPCAP}$ may be explained by instances where the assumed intrinsic colors are
inappropriate for those stellar types.  The NIR colors systematically overestimate the extinction, particularly at high extinction,
which could indicate that the assumed $(H-K_s)_0$ is offset from the actual mean color of the stars or that the assumed
extinction law is incorrect.  
The dependence of $\Delta A(K_s)$ on the stellar parameters strongly suggests
that the offset extinctions are due to the fact that the assumed $(H-K_s)_0$ color is not applicable to all stellar types.
As shown by \cite{bessell1989} and \cite{lancon2000}, $(H-K_s)_0$ is very sensitive to gravity and metallicity,
making the assumption of a single intrinsic color for stars of unknown stellar properties inappropriate.
For the stars in our sample, this color appears to be increasingly inapplicable as one moves up the RGB, to lower $\log{g}$
and $T_{\rm eff}$, with median $A(K_s)$ discrepancies of 0.15 mag for stars with $\log{g} \lesssim 1.75$ and $T_{\rm eff} \lesssim 4000$~K
(corresponding to the common late K giants).  

In contrast, the individual RJCE extinction estimates appear to be largely accurate for nearly all stars, 
with the exception of the coolest RGB stars (M type).  
The small consistent offset observed ($\lesssim$0.03--0.05 mag) is almost entirely independent of the stellar properties,
indicating that a change in the assumed intrinsic color of $\sim$0.03 mag would remove the discrepancy.
We conclude that the RJCE method is the most robust method for tracing interstellar extinction 
spanning the $T_{\rm eff}$, $\log{g}$, and [M/H] stellar parameter space studied. However, we do note that  if one averages the stellar RJCE extinction values (not just the APOGEE sample, but all bright 2MASS stars) over an area
 comparable to the N12 and G12 map pixels, the dispersion is comparable to the scatter in Figure 4, strongly suggesting that unresolved differential extinction  also has an impact on those map values. 




\section{Comparison to 3D Maps and Models} \label{sec:compare3d}

\subsection{3D maps}

For comparison to 3D extinction distributions, we use the following data:
\begin{itemize}
\item \citet{marshall2006}  used the stellar
population synthesis model of Besan\c{c}on (\citealt{robin2003}), together with the 2MASS data set, to map the 3D extinction for
$|l| \le 90^\circ$ and $|b| \le 10^\circ$.  Assuming a distance vs. color relation (after removing the M-dwarf
foreground population), they compared the observed stellar colors to the synthetic ones for each line of sight and
attributed the resulting reddening to specific distances according to the model. 
This study is somewhat limited, due to the confusion limit of 2MASS ($\sim$3\arcsec pixels) in the crowded Galactic bulge region and by the sensitivity of 2MASS
 in highly extinguished regions. The spatial resolution of the map is 15\arcmin.
\item \citet{chen2012} combined the {\it Spitzer}-IRAC GLIMPSE-II data with the VVV data, along with an improved version
of the Besan\c{c}on model (\citealt{robin2012}), to map the inner bulge region ($|l| \le 10^\circ$ and $|b| \ge 2^\circ$) in 3D. An extension of this map for the entire
VVV bulge area ($\rm -10^\circ \le l \le 10^\circ$, $\rm -10^\circ \le b \le 5^\circ$) has been provided by \citet{schultheis2014}.
They used  an improved color-temperature relation for M giants (using the Padova isochrones; \citealt{girardi2010}). In addition, they  fit the full colour-magnitude diagram including dwarf stars, to the synthetic CMD, to derive the
3D-extinction.

Their spatial resolution is also 15\arcmin. 
\item \citet{drimmel2003} present a Galactic-scale 3D model of Galactic extinction based on the dust distribution model of \citet{drimmel2001}, 
which is fitted to the far- and near-IR data from the COBE/DIRBE instrument. The size of the COBE pixels are approximately 
$21\arcmin \times 21\arcmin$. 
\end{itemize}

\noindent Again, if necessary, we rescale all the maps to $A(K_s)$ using the \citet{nishiyama2009} bulge extinction law. 


We queried these three maps for the 
extinction at the position $(l,b)$ of each of our APOGEE bulge stars, within a FOV of 0.25~deg$^2$ and in distance bins of 1~kpc.
This FOV was chosen to contain a sufficient number of APOGEE stars with accurate stellar parameters
at each spatial position. Because the
extinction maps have different spatial resolutions, we took the median
value around the center position of each 0.25~deg$^2$ field. 
Figure~\ref{los1} shows the comparison of the 3D extinction along different lines of sight, where we compare the literature extinction values (colored lines) 
with the isochrone-derived $A(K_s)$ values for the APOGEE sources (black points). 
We indicate also the errors in $A(K_s)$ and distance. 
Note that, in contrast to \citet{drimmel2003}, the 3D maps of \citet{marshall2006} and \citet{schultheis2014} are limited to a heliocentric distance of $<$10~kpc.

\subsection{Results and Discussion} \label{sec:discussion3d}

Figure~\ref{los1} demonstrates  that none of the different 3D models reproduce the global $A(K_s)_{\rm ASPCAP}$ vs. distance relations along all available lines of sight. 
The \citet{drimmel2003} 3D Galactic dust distribution model has, in general  the largest systematic deviation from our observed $A(K_s)$ vs. distance relations.
Drimmel et al.\,showed (see their Figure~10) that in the inner disk of the Galaxy, 
their $K_s$-band extinction can deviate on the order of 20\% compared to extinction derived from 2MASS color-magnitude diagrams. 
  

Along most of the lines of sight, we lack APOGEE stars at distances closer to the Sun than $\sim$4~kpc. 
Nevertheless, we note a number of trends.
\begin{itemize}
\item For most of the lines of sight, we confirm the steep rise in  $A(K_s)$, with a flattening occurring at $\sim$4--6~kpc, predicted by all of the 3D distributions. 
In the highly extinguished regions $(l,b) = (+8^\circ,0^\circ)$, $(+2^\circ,0^\circ)$, $(-2^\circ,0^\circ)$, and $(0^\circ,0^\circ)$, however, this flattening is not predicted by the
models. Unfortunately, we do not currently have a sufficient number of data points to confirm this.

\item The Marshall et al. 3D model appears to best represent the increase of $A(K_s)$ with distance for smaller distances ($d < 4$~kpc),
while Schultheis et al.\, most reliably predicts extinction for larger distances ($d > 4$~kpc). 
The Marshall et al. map is confusion limited by 2MASS, which produces the sudden decrease in $A(K_s)$ at around 8\,kpc. 

\item The Drimmel et al. map systematically overestimates extinction in the fields 
$(l,b) = (+4^\circ,+2^\circ)$, $(-2^\circ,+1^\circ)$, $(+1^\circ,-4^\circ)$,  and $(0^\circ,-5^\circ)$,
and underestimates for the fields  $(l,b) = (+2^\circ,-1.5^\circ)$, $(+1^\circ,+1.5^\circ)$, $(0^\circ,+4.5^\circ)$,  and $(l,b) = (+2^\circ,+4^\circ)$.

\item In the innermost bulge region ($|l| \le 2^\circ, |b| \le 1^\circ$), despite low number statistics, a steep rise in $A(K_s)$ is detected within 3~kpc of the Sun
that is {\em{only}} predicted by the Drimmel et al. model.

\item  The APOGEE stars in the fields $(l,b) = (0^\circ,-5^\circ)$ and $(+1^\circ,-4.0^\circ)$ span the full range of distance from $\le$\,2--10~kpc. 
These low-extinction fields follow the $A(K_s)$ vs. distance relation predicted by both \citet{marshall2006} and \citet{schultheis2014}, 
while \citet{drimmel2003} overestimates the extinction, particularly where $d \gtrsim 5$~kpc. 
Due to the low extinction, however, small spatial variations in $A(K_s)$ cannot be traced with this data set
--- optical data, with its greater extinction susceptibility, is necessary.

\end{itemize}


With additional data from APOGEE and APOGEE-2 in the coming years, we will be able to trace
the distance vs.\,$A(K_s)$ behavior towards the full Galactic bulge, making this sample an ideal tool with which to systematically and quantitatively compare
spectroscopically-derived values with existing 3D interstellar extinction maps, and to better understand the dust structure.


\begin{figure}[!htbp]
   \includegraphics[width=9.0cm]{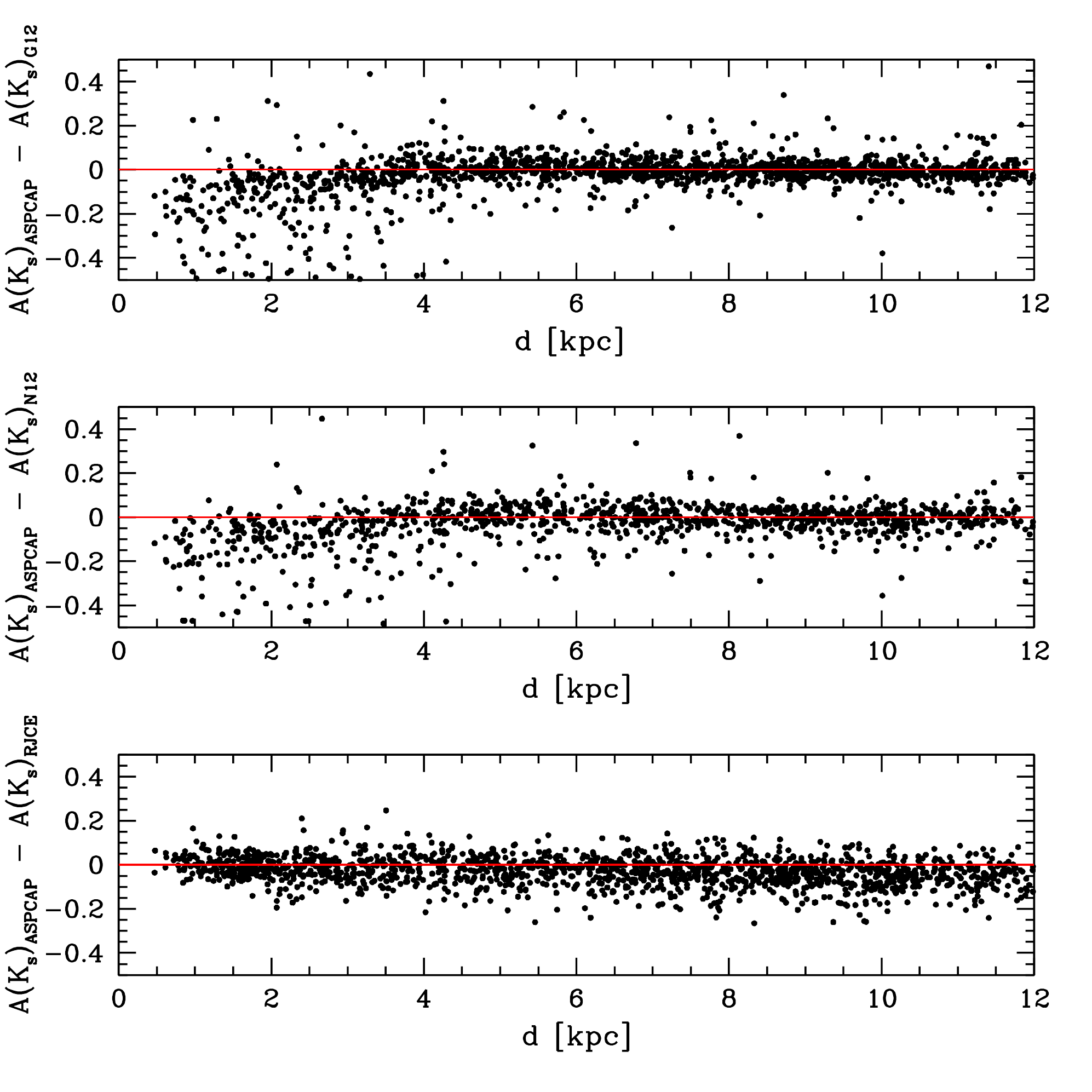}
\caption{Difference of our derived extinction, $A(K_s)_{\rm ASPCAP}$; to those of G12 (upper panel), N12 (middle), and RJCE (lower) extinction values,  as a function of distance.}
\vspace*{0.6cm}
\label{DeltaAkvsdist}
\end{figure}

\begin{figure}[!htbp]
   \includegraphics[width=8.0cm]{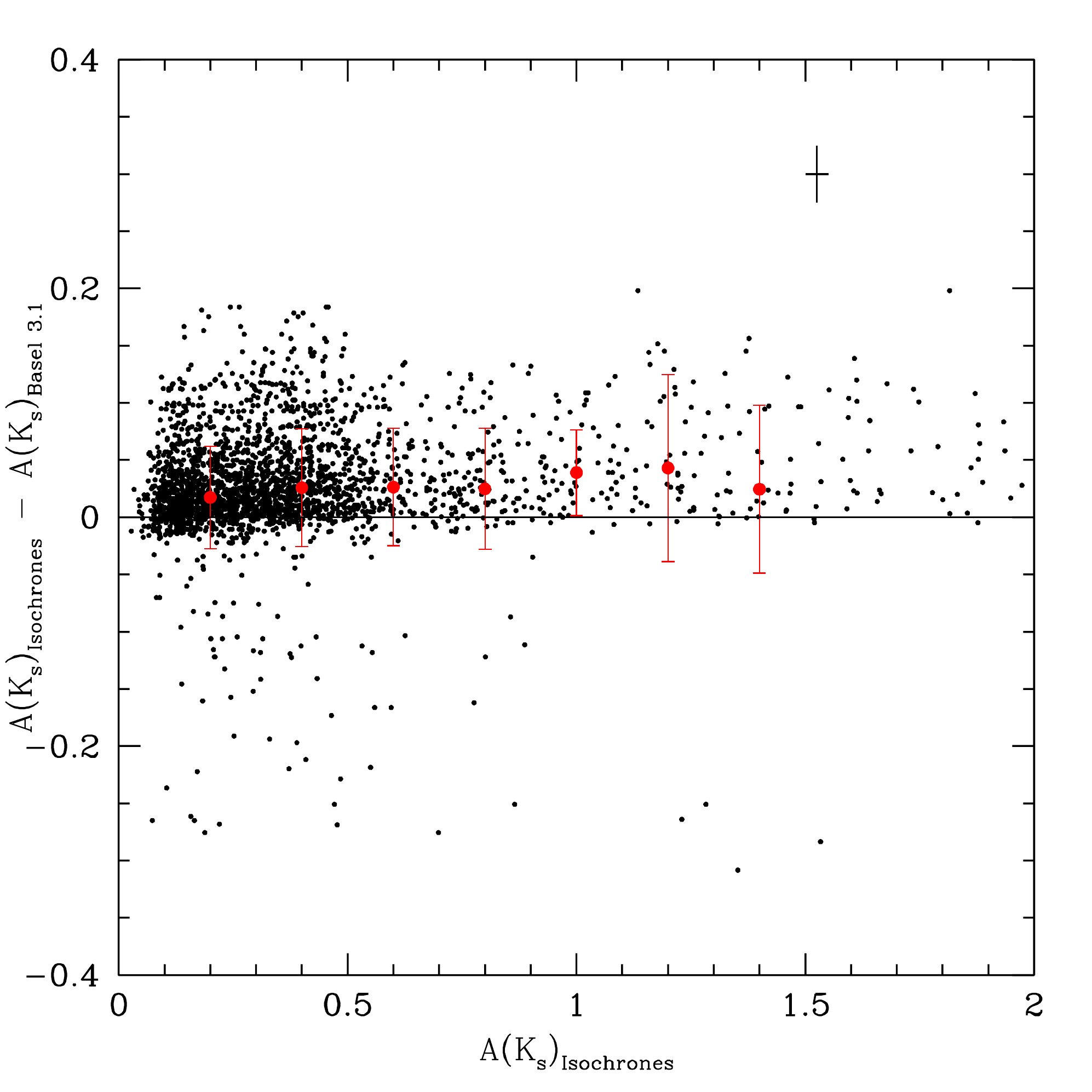}
\caption{Extinction derived from Padova isochrones compared to those obtained using the Basel3.1 stellar library.
The red points indicate the median value of the differences. The typical uncertainty bar is shown in the top right corner}
\label{Akstellib}
\end{figure}

\begin{figure}[!htbp]
   \includegraphics[width=8.0cm]{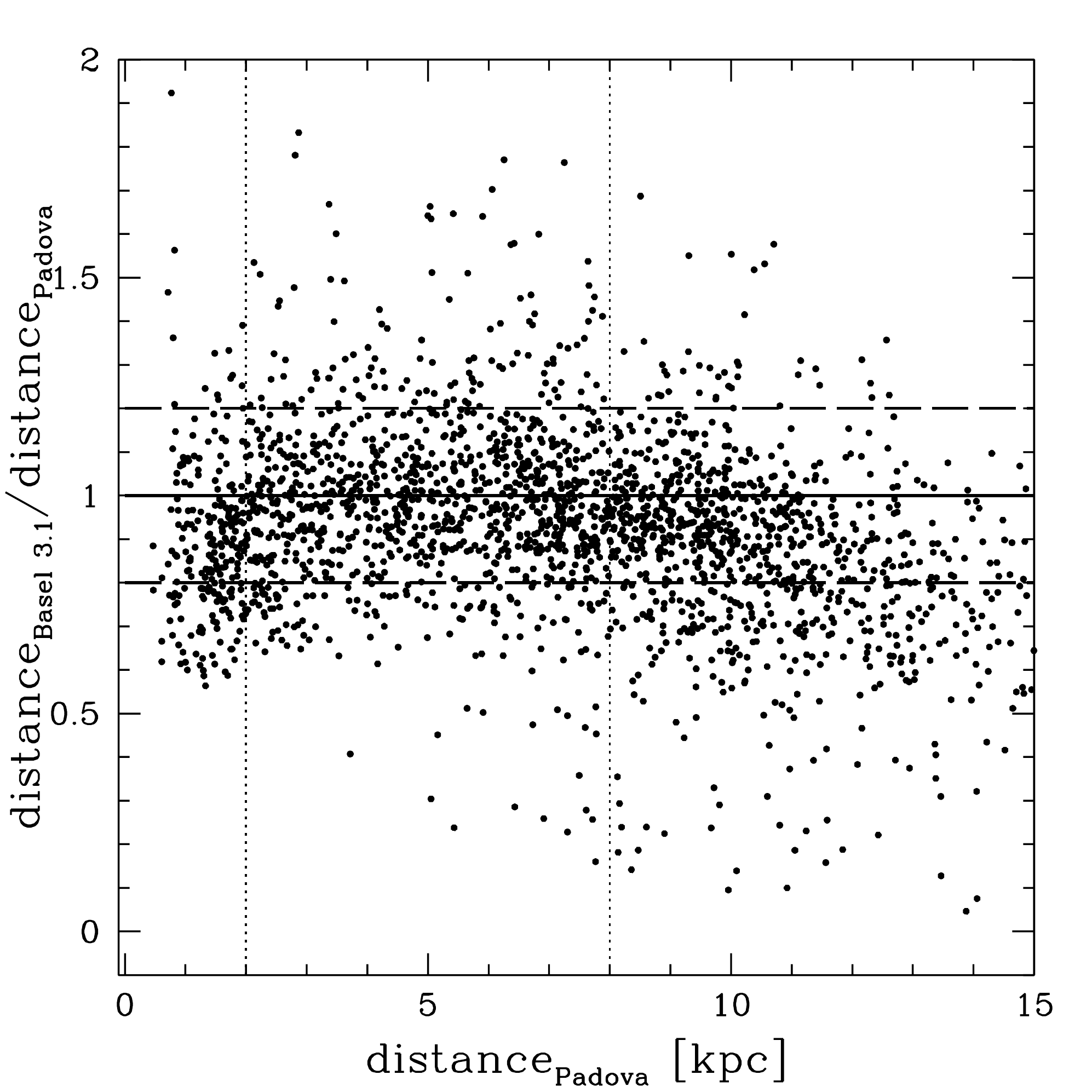}
\caption{Ratio of distances derived from the Padova isochrones to the Basel3.1
 stellar library, as a function of distance.
The dashed horizontal lines indicate $\pm 20\%$ error.}
\label{diststellib}
\end{figure}

\begin{figure*}[!htbp]
\begin{center}
   \includegraphics[width=0.3\textwidth]{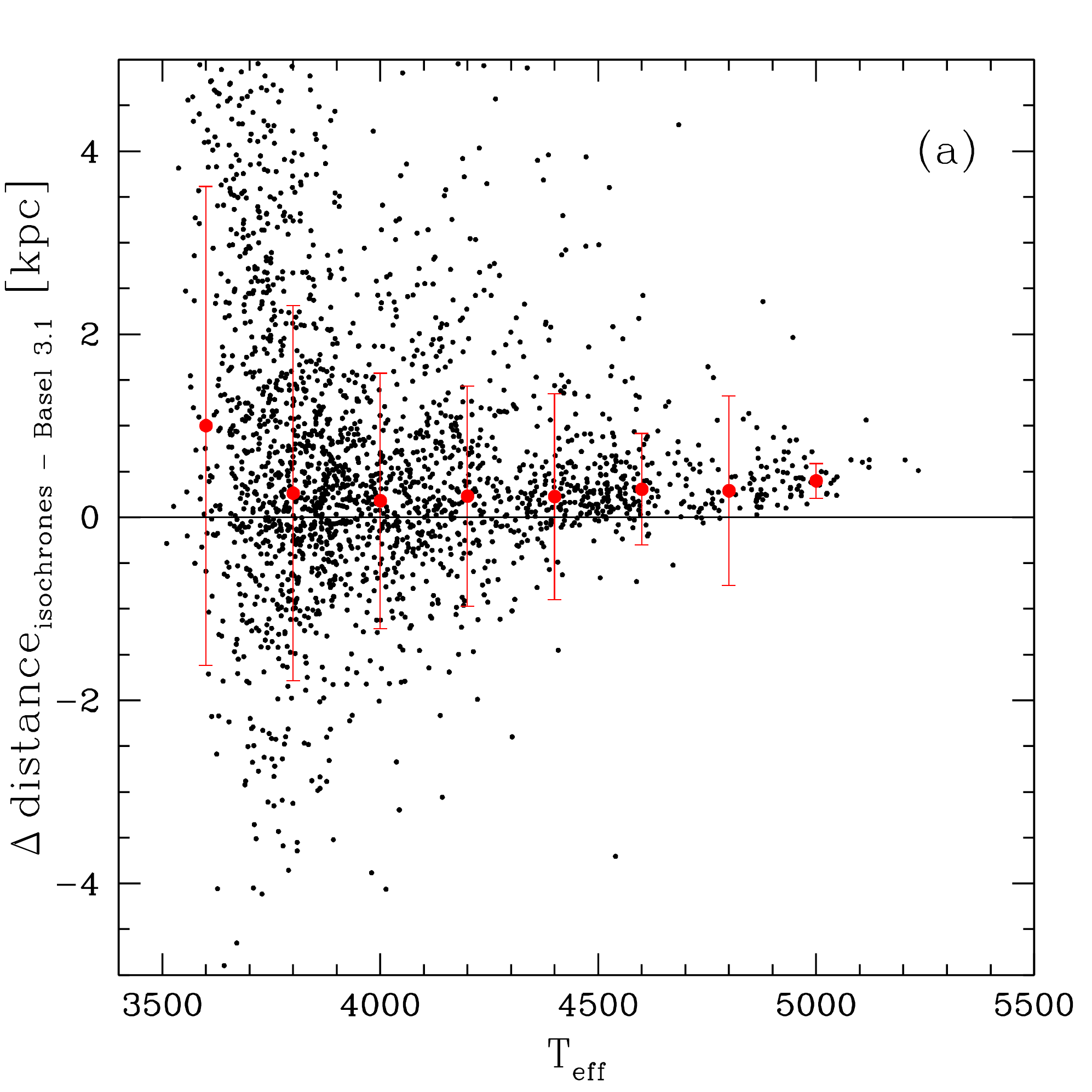}
   \includegraphics[width=0.3\textwidth]{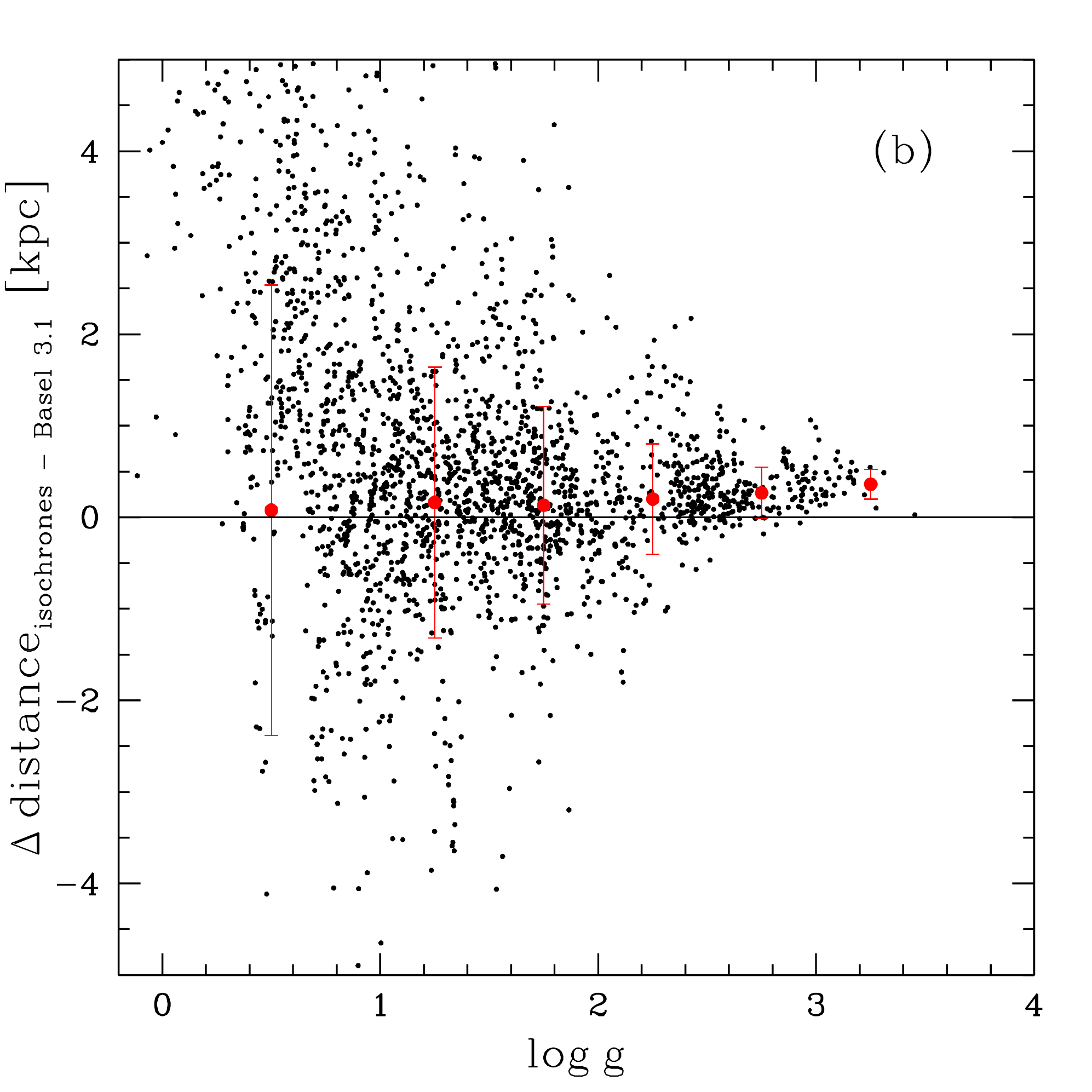}
   \includegraphics[width=0.3\textwidth]{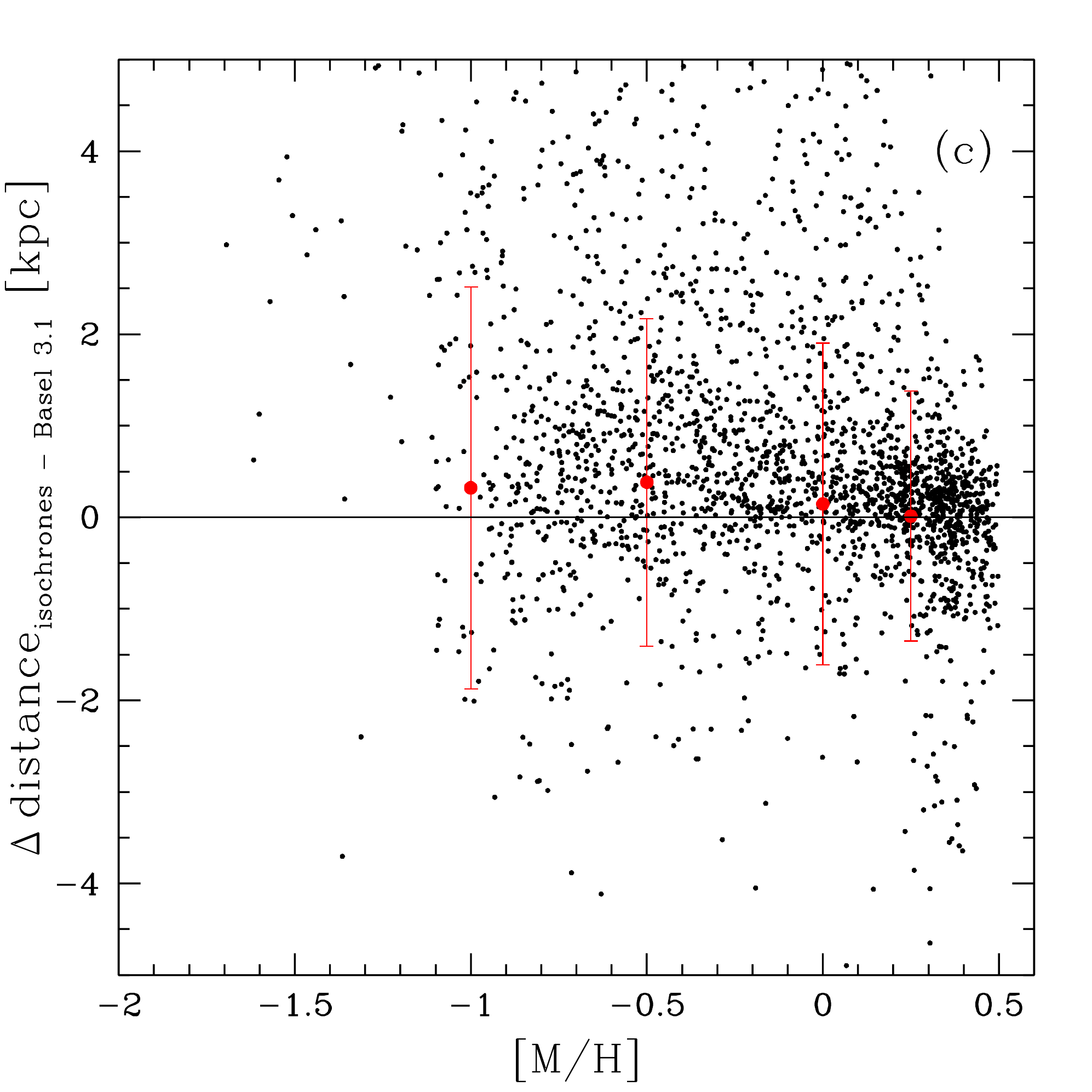}
\caption{Difference in distances between Padova and Basel3.1, as a function of (a) $T_{\rm eff}$, (b) $\log{g}$, and (c) [M/H].}
\vspace*{0.4cm}  
\label{distparamstellib}
\end{center}
\end{figure*}

\section{Additional Discussion} \label{sec:discussion}

\subsection{2D vs. 3D Extinction Maps} \label{sec:2d-vs-3d-maps}

What are the biases introduced by adopting a 2D extinction map over a more detailed  3D distribution?
What are the typical distances probed by 2D bulge extinction maps?
In Figure~\ref{DeltaAkvsdist}, we present the difference in $A(K_s)$ as a function of stellar distance for the 2D maps N12 and G12. 
Both 2D maps agree remarkably well --- with each other and with the ASPCAP extinction ---
for distances larger than $\sim$3--4\,kpc, while for the shorter distances they systematically overestimate
extinction by $\sim$0.1--0.2 mag ($\sim$0.8--1.6 mag in $V$-band). 
This result suggests that for sources known to be in the Galactic bulge, or at least beyond $\sim$4~kpc from the Sun, 
these 2D extinction maps can be applied without significant systematic offsets. 
The RJCE color excess method, which is in theory distance-independent (if not entirely in practice, due to the need for quality stellar photometry),
also traces the extinction well for shorter distances, provided the star is not too metal-poor (Section~\ref{sec:discussion2d}).  
Thus, the primary potential bias in the application of 2D maps is for stars along the line of sight towards the bulge, but not established to be in it.

\subsection{Dependence on Extinction Law} \label{sec:extlaw}

The derived distances also depend on the extinction law adopted to convert between the reddening
measured from the isochrone-matching and the final extinction used in the distance calculation.  
We assume here the extinction law derived
for the Galactic bulge by \citet{nishiyama2009}, with $A(K_s) = 0.526 \times E(J-K_s)$. 
However, studies have found spatial variations in the extinction-law coefficients even within the bulge 
(e.g., \citealt{udalski2003}, \citealt{gosling2009}; \citealt{chen2012}; \citealt{nataf2013}),
which may be due to changes in the dust size distribution (e.g., \citealt{draine2003}; \citealt{chapman2009}; \citealt{zasowski2009}). 
We tested the impact of adopting a different extinction law  and found that, compared to our distance errors from other sources,
any shift (even with the standard \citealt{cardelli1989} law) is sufficiently negligible  and does not affect our conclusions here. A difference in $\sim$10\% in the
extinction law produces a typical difference in the distance estimate of
about 5\%, which is smaller than our derived errors.

\subsection{Differences Between Stellar Libraries} \label{stellib}
The stellar atmosphere models used to produce the isochrones also affect
the final extinction values and distances derived for the stars.
Here we compare the distances used in this analysis, calculated with the Padova isochrones \citep{girardi2010},
with those calculated using an identical method but adopting the Basel3.1 model library (\citealt{lejeune1997}).
We demonstrate the importance of being aware that model libraries do differ, and that the differences can complicate the comparison of
results that do not use the same library.

 The Basel3.1 library is a semi-empirical, stellar atmospheric library based on the preceding Basel2.2 generation of models (\citealt{lejeune1997})
 and extended to non-solar metallicities by \citet{westera2002}. 
 The Kurucz theoretical spectra (\citealt{kurucz1979}) have been modified to fit broad-band photometry  (\citealt{Buser1992}). 
As shown by \citet{schultheis2006}, the Basel3.1 library produces reliable
color estimates for $T_{\rm eff} > 4000$~K but shows significant offsets for cooler stars.
Therefore, we have used complementary NextGen models from PHOENIX stellar atmosphere models (\citealt{hauschildt1997}) for stars with $T_{\rm eff} < 4000$~K.  
As the PHOENIX models use a direct opacity sampling, including over 500 million lines of atomic and molecular species, 
they yield a more realistic description of the M-star population.  
This composite stellar library is also the one used in the Besan\c{c}on stellar population synthesis model (\citealt{robin2012}).

The effect on the derived $A(K_s)$ of using different stellar libraries is presented in Figure~\ref{Akstellib}. 
The Basel3.1 $A(K_s)$ values are systematically smaller than the Padova ones, 
which indicates that the model intrinsic colors are slightly redder than those in the Padova library. 
 However, the effect is typically smaller than $0.05$ mag in $A(K_s)$.


Figure \ref{diststellib} shows the comparison in the distances, $d$, derived using the Padova
and the Basel3.1 isochrones.  Some systematic difference
is apparent,  especially for $d \lesssim 2$~kpc and $d \gtrsim 10$~kpc, where the mean Basel3.1 distances are systematically smaller. The typical
dispersion is about 20\% (indicated by the dashed lines), 
which is roughly comparable to the intrinsic error of our method (see Figure~\ref{distancevserrordist}).

Figure~\ref{distparamstellib} traces the distance differences as functions of the spectroscopic stellar parameters.
Figure~\ref{distparamstellib}a reveals a dramatically larger dispersion for stars with $T_{\rm eff} < 3800$~K, 
which suggests important systematic differences in the stellar atmosphere models between Padova and Basel3.1. 
(This is close to --- but not exactly --- where the Kurucz and PHOENIX model atmospheres are merged in the Basel3.1 library.)
Similarly, in Figure\,\ref{distparamstellib}b, a larger
 spread in $\Delta d$ for stars with $\log{g} < 2$ is found, along with consistently larger distances where $\log{g} < 0.4$.
There are also  larger discrepancies when going to lower metallicities (${\rm [M/H]} \lesssim +0.25$; Figure~\ref{distparamstellib}c). 
Thus, the use of different stellar libraries can produce systematic differences in calculated distances, 
most significantly for cool, metal-poor M giants.  As these stars are often the most distant objects  in a magnitude-limited sample of giants, such as  APOGEE's,
care must be taken in the interpretation of results apparently coming from the distant zones probed by these stars.

\section{Conclusions}
%


Data from the APOGEE survey, which includes stellar parameters ($T_{\rm eff}$, $\log{g}$, [M/H]) as well as spectra, serve as useful tools to trace interstellar extinction
in the most heavily extinguished parts of the Milky Way --- the bulge and inner disk.  In this paper, we matched the fundamental parameters of 
2433 bulge giant stars to model isochrones and derived their extinctions and distances (Sections~\ref{sec:sample} and \ref{sec:extdistances}).  
We compared these spectroscopically-derived extinctions to a variety of theoretical and photometrically-derived extinction maps and models
(Sections~\ref{sec:compare2d} and \ref{sec:compare3d}).

Individual stellar extinction estimates derived from long baseline, near- to mid-IR colors (the RJCE method; \citealt{majewski2011})
are the most reliable predictors of extinction towards our sample.  Extinctions based on near-IR colors only appear accurate for a much narrower range of 
spectral types, which then requires foreknowledge of (or assumptions about) the stars under consideration before calculating the extinctions.
The bulge extinction maps of \citet{nidever2012} and \citet{gonzalez2012} perform similarly well in tracing the extinction of stars actually residing 
in the bulge, but unsurprisingly, overpredict the extinction for foreground sources.  For the \citet{nidever2012} and \citet{gonzalez2012}  maps,  ``foreground'' means
stars closer to the Sun than $\sim$3--4~kpc.
 
We also use our stellar distance estimates to assess three 3D extinction distributions: 
the model of \citet{drimmel2003} and the maps of \citet{marshall2006} and \citet{schultheis2014}.
We confirm the steep increase in extinction in the first few kpc, and the flattening
of the extinction at about 4~kpc from the Sun. However, none of the 3D maps agree for all of the APOGEE lines of sight, 
which demonstrates that there remains significant room for improvement in our 
knowledge of the 3D bulge extinction distribution.

Finally, we examine additional sources of uncertainty in our comparisons: variations in the adopted extinction law and variations in the isochrones
from different stellar atmospheric models (Section~\ref{sec:discussion}).  Uncertainties due to potential extinction law variations are 
small compared to uncertainties from the stellar parameters and distance calculations themselves.  However, discrepancies in 
derived extinctions and distances can be substantial if different atmospheric models are adopted, 
particularly for the coolest (and typically most distant) giants.  This caveat should be considered by anyone wishing to 
compare or combine results that use different atmospheric libraries.

\begin{acknowledgements}
We want to thank the referee for his/her helpful comments.
GZ has been supported by an NSF Astronomy \& Astrophysics Postdoctoral Fellowship under Award No.\ AST-1203017. TCB acknowledges partial support from support from grant PHY 08-22648: Physics Frontier Center / Joint Institute for Nuclear Astrophysics (JINA), awarded by the U.S. National Science Foundation. MS has been supported by the ANR-12-BS05-0015-01.

Funding for SDSS-III has been provided by the Alfred P. Sloan Foundation, the Participating Institutions, the National Science Foundation, and the U.S. Department of Energy Office of Science. The SDSS-III web site is http://www.sdss3.org/.

SDSS-III is managed by the Astrophysical Research Consortium for the Participating Institutions of the SDSS-III Collaboration including the University of Arizona, the Brazilian Participation Group, Brookhaven National Laboratory, Carnegie Mellon University, University of Florida, the French Participation Group, the German Participation Group, Harvard University, the Instituto de Astrofisica de Canarias, the Michigan State/Notre Dame/JINA Participation Group, Johns Hopkins University, Lawrence Berkeley National Laboratory, Max Planck Institute for Astrophysics, Max Planck Institute for Extraterrestrial Physics, New Mexico State University, New York University, The Ohio State University, Pennsylvania State University, University of Portsmouth, Princeton University, the Spanish Participation Group, University of Tokyo, University of Utah, Vanderbilt University, University of Virginia, University of Washington, and Yale University. 
\end{acknowledgements}

\bibliographystyle{aa}
\bibliography{apogee_3dextinction_revised_accepted}

\end{document}